\documentclass[11pt]{article}
\usepackage{moriond}
\usepackage{amsfonts}
\usepackage{epsfig}
\newsymbol\gtrsim 1326

\def\MET{\mbox{${\hbox{$E$\kern-0.6em\lower-.1ex\hbox{/}}}_T$}} 
\def\D0{D\O}                            

\begin{document}
\title{EXTRA DIMENSIONS AND MORE...\footnote{Full version of this talk is available from {\tt http://hep.brown.edu/users/Greg/talks/Moriond01.pdf}}}

\author{GREG LANDSBERG}

\address{Brown University, Department of Physics, 182 Hope St., Providence, RI 02912, USA\\E-mail: landsberg@hep.brown.edu}

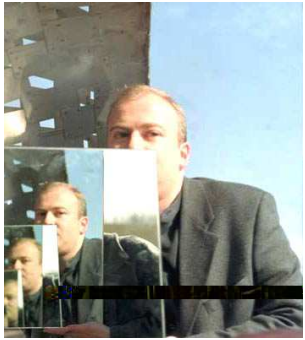
\maketitle
\vspace*{-0.4in}
\abstracts{
One of the most stimulating recent ideas in particle physics involves a possibility that our universe has additional compactified spatial dimensions, perhaps as large as 1~mm. In this review, we discuss the results of recent experimental searches for such large extra dimensions, as well as new developments in this field.}
\vspace*{-0.1in}

The possibility that the universe has more than three spatial dimensions has been discussed since it was first suggested by Riemann. String theory would have us believe that there could be up to $n=7$ additional dimensions, compactified at distances of the order of $10^{-32}$~m. In a new model,\cite{ADD} inspired by string theory, several of the compactified extra dimensions are suggested to be as large as 1~mm. These large extra dimensions (LED) are introduced to solve the hierarchy problem of the standard model (SM) by lowering the Planck scale ($M_{\rm Pl}$) to an energy range of a TeV. We refer to this effective Planck scale as $M_S$.

Since Newton's law of gravity in the presence of compactified extra dimensions is modified for interaction distances below the size of the LED, current gravitational observations rule out the possibility of only a single LED. Recent results from gravity experiments at submillimeter distances ($M_S > 1.9$~TeV for $n=2$),\cite{tabletop} as well as astrophysical constraints from supernova cooling and cosmic diffuse gamma radiation ($M_S \gtrsim $~30--100~TeV for $n=2$),\cite{cosmology} indicate that the case for $n = 2$ is also likely ruled. For $n \ge 3$, the sizes of the LED become microscopic, and therefore elude the reach of direct gravitational measurements or current astrophysical constraints. However, high energy colliders, capable of probing very short distances, can provide crucial tests of the LED hypothesis, in which effects of gravity are enhanced at high energies due to accessibility of numerous excited graviton states that become wrapped around the compactified dimensions.

LED phenomenology at colliders has already been studied in detail.\cite{GRW,HLZ,Peskin,Hewett} One of the primary observable effects is an apparent non-conservation of momentum caused by the direct emission of  gravitons that leave the three flat spatial dimensions. A typical signature is the production of a single jet or a vector boson at large transverse momentum. The other observable effect is the anomalous production of fermion-antifermion or diboson pairs with large invariant mass stemming from the coupling to virtual gravitons. Direct graviton emission is expected to be suppressed by a factor $(M_S)^{n+2}$, while virtual graviton effects depend only weakly on the number of extra dimensions.\cite{GRW,HLZ,Hewett} Virtual graviton production therefore offers a potentially more sensitive way to search for manifestations of LED.\footnote{Strictly speaking, virtual graviton effects are sensitive to the ultraviolet cutoff required to keep the divergent sum over the graviton modes finite.\protect\cite{GRW,HLZ,Hewett} This cutoff is expected to be of the order of the effective Planck scale. Dependence on the value of the cutoff is discussed in, e.g., Refs.\protect\cite{Shrock,KCGL}}

The effects of direct graviton emission, including production of single photons or $Z$'s, have been sought at LEP.\cite{common,direct,L3ZZold} The following signatures were used: $\gamma\MET$ or $Z(\to jj)\MET$, where \MET\ is the missing transverse energy in the detector, and $j$ stands for jet. The negative results of these searches can be expressed in terms of limits on the effective Planck scale, as summarized in Table~\ref{direct}. The CDF and D\O\ Collaborations at the Fermilab Tevatron are also looking for direct graviton emission in the ``monojet'' ($j\MET$) channel, which is quite challenging due to large instrumental background from jet mismeasurement and cosmic rays. Although no results have as yet been reported, the sensitivity of these searches is expected to be similar to those at LEP.

\begin{table}[hbt]
\begin{center}
\vspace*{-0.1in}
\caption{Lower limits at the 95\% CL on the effective Planck scale in TeV, from searches for direct graviton production at LEP. Limits from $\sqrt{s} > 200$~GeV data are shown in normal font; limits from 189~GeV data are in {\it italics\/}; limits from 184~GeV are in {\bf bold} script.}
\label{direct}
\medskip
\begin{tabular}{|l|ccccc|ccccc|}
\hline
{\footnotesize\rm Experiment}  & \multicolumn{5}{|c|}{$e^+e^- \to \gamma G_{\rm KK}$} &  
              \multicolumn{5}{c|}{$e^+e^- \to Z G_{\rm KK}$} \\
\cline{2-6}\cline{7-11}
		& $n$=2 & $n$=3 & $n$=4 & $n$=5 & $n$=6 & 
              $n$=2 & $n$=3 & $n$=4 & $n$=5 & $n$=6 \\
\hline
ALEPH & 1.28 & 0.97 & 0.78 & 0.66 & 0.57 & \bf 0.35 &  \bf 0.22 & \bf 0.17 & \bf 0.14 & \bf 0.12 \\
\hline
DELPHI& 1.38 & 1.02 & 0.84 & 0.68 & 0.58 & N/A  & N/A  & N/A  & N/A  & N/A  \\
\hline
L3	& \it 1.02 & \it 0.81 & \it 0.67 & \it 0.58 & \it 0.51 & \it 0.60 & \it 0.38 & \it 0.29 & \it 0.24 & \it 0.21 \\
\hline
OPAL	& \it 1.09 & \it 0.86 & \it 0.71 & \it 0.61 & \it 0.53 & N/A  & N/A  & N/A  & N/A  & N/A  \\
\hline
\end{tabular}
\end{center}
\end{table}

While the formalism for calculating direct graviton emission is well established, different formalisms have been used to describe virtual graviton effects.\cite{GRW,HLZ,Hewett} Since difermion or diboson production via virtual graviton exchange can interfere with SM production of the same final-state particles, the cross section in the presence of LED is given by:~\cite{GRW,HLZ,Hewett} $\sigma = \sigma_{\rm SM} + \sigma_{\rm int}\eta_G + \sigma_G\;\eta_G^2,$ where $\sigma_{\rm SM}$, $\sigma_{\rm int}$, and $\sigma_G$ denote the SM, interference, and graviton terms, and the effects of LED are parameterized via a single variable $\eta_G = {\cal F}/M_S^4$, where ${\cal F}$ is a dimensionless parameter of order unity. Several definitions exist for ${\cal F}$:
\vspace*{-0.1in}
\begin{eqnarray}
	{\cal F} & = & 1, \mbox{~(GRW~\cite{GRW});} \nonumber\\
	{\cal F} & = & \left\{ \begin{array}{ll} 
         \log\left( \frac{M_S^2}{M^2} \right), & n = 2 \\
	   \frac{2}{n-2}, & n > 2
	   \end{array} \right. , \mbox{~(HLZ~\cite{HLZ});} \nonumber\\
	{\cal F} & = & \frac{2\lambda}{\pi} = \pm\frac{2}{\pi}, \mbox{~(Hewett~\cite{Hewett}).} \nonumber
\end{eqnarray}
Here, $\lambda$ is a dimensionless parameter of order unity, conventionally set to be either $+1$ or $-1$ in calculations within Hewett's formalism. Only the HLZ formalism has ${\cal F}$ depending explicitly on $n$. Because different experiments have set limits on virtual graviton exchange using different formalisms, it is worthwhile to specify relationship between the three definitions of effective Planck scale, referred to as $\Lambda_T$,\cite{GRW} $M_S({\rm Hewett})$, and $M_S({\rm HLZ})$: $M_S({\rm Hewett})\left|_{\lambda=+1}\right.  = 
  \sqrt[4]{\frac{2}{\pi}} M_S({\rm HLZ})\left|_{n=4}\right.\!\!;
	\Lambda_T =  M_S({\rm HLZ})\left|_{n=4}\right..$
All the limits on the effective Planck scale are given at 95\% CL, and expressed in terms of $M_S({\rm Hewett})$.

Among the many difermion and diboson final states tested for presence of virtual graviton effects at LEP,\cite{common,L3ZZold,virtual,L3ZZ,OPALZZ} the most sensitive channels involve the dielectron (both Drell-Yan and Bhabha scattering) and diphoton processes.\footnote{
Preliminary results from L3 (circa summer 2000, not updated since) at $\sqrt{s} > 200$~GeV indicate that the best sensitivity is found in the $ZZ$ channel,\protect\cite{L3ZZ} but details of the experimental analysis are not yet available. This is different from an earlier L3 publication,\protect\cite{L3ZZold} where the sensitivity in the $ZZ$ channel at $\sqrt{s} = 189$~GeV was significantly lower than that in the $\gamma\gamma$ channel. Recent OPAL results in the $ZZ$ channel at the highest LEP energies,\protect\cite{OPALZZ} also support the same conclusion. It may therefore be prudent to await final results from L3 on this issue.} 
None of the experiments see any significant deviation from the SM. This is translated into the limits on $M_S({\rm Hewett})$, listed in Table~\ref{virtual}. They are of the order of 1~TeV for both signs of interference term.

\begin{table}[hbt]
\begin{center}
\vspace*{-0.1in}
\caption{Lower limits at the 95\% CL on the effective Planck scale, $M_S({\rm Hewett})$, in TeV, from searches for virtual graviton effect at LEP. Upper (lower) rows correspond to $\lambda=+1$ ($\lambda = -1$). The ALEPH Collaboration used the GRW formalism for their analysis, so their limits were translated into Hewett's formalism. The L3 Collaboration used formalism~\protect\cite{AD} for diboson production in which the sign of $\lambda$ is reversed compared to Hewett. To correct for that, we reverse the sign of $\lambda$ when quoting the L3 limits in the $\gamma\gamma$, $WW$, and $ZZ$ channels. Combined L3 limits are nevertheless affected by the mixture of two signs of $\lambda$ in difermion and diboson channels. (See also footnote on previous page for a discussion of $ZZ$ results.) Limits from $\sqrt{s} > 200$~GeV data are shown in normal font; limits from 189~GeV data are in {\it italics\/}; limits from 184~GeV are in {\bf bold} script.}
\label{virtual}
\medskip
\begin{tabular}{|l|cccc|c|ccc|c|}
\hline
{\footnotesize\rm Experiment}  & $e^+e^-$ & $\mu^+\mu^-$ & $\tau^+\tau^-$ & $q\bar q \quad (b\bar b)$ & $f\bar f$ & $\gamma\gamma$ & $WW$ & $ZZ$ & {\footnotesize\rm Combined} \\
\hline
ALEPH & 0.81 & 0.67 & 0.62 & 0.57 (0.49) & 0.84 & 0.82 & N/A  & N/A  & \it 1.00 \\
	& 1.04 & 0.65 & 0.60 & 0.53 (0.49) & 1.05 & 0.81 & N/A  & N/A  & \it 0.75 \\
\hline
DELPHI& N/A  & 0.73 & 0.65 & N/A  (N/A)  & 0.76 & 0.77 & N/A  & N/A  & N/A \\
	& N/A  & 0.59 & 0.56 & N/A  (N/A)  & 0.60 & 0.70 & N/A  & N/A  & N/A \\
\hline
L3	& 0.99 & \it 0.69 & \it 0.54 & {\bf 0.49} (N/A)  & \it 1.00 & 0.99 & \it 0.68 & 1.2  & 1.3 \\
	& 0.91 & \it 0.56 & \it 0.58 & {\bf 0.49} (N/A)  & \it 0.84 & 0.84 & \it 0.79 & 1.2  & 1.2 \\
\hline
OPAL	& 1.00  & \multicolumn{2}{c}{0.66}  & N/A  (N/A)  & 0.66 & 0.83 & N/A  & 0.74 & 1.03 \\
	& 1.15  & \multicolumn{2}{c}{0.62}  & N/A  (N/A)  & 0.62 & 0.89 & N/A  & 0.63 & 1.17 \\
\hline
\end{tabular}
\end{center}
\end{table}

Virtual graviton effects have also been sought at HERA in the $t$-channel of $e^\pm p \to e^\pm p$ scattering, similar to Bhabha scattering at LEP.\cite{GRW,Hewett} A search carried out by the H1 Collaboration~\cite{H1} with 82~pb$^{-1}$ of $e^+p$ and 15~pb$^{-1}$ of $e^-p$ data, have set limits on $M_S$ between 0.5 and 0.8~TeV. Similar limits were recently reported by ZEUS~\cite{ZEUS} (see Table~\ref{H1}). Although these limits are somewhat inferior to those from LEP, the ultimate sensitivity of HERA at the end of the next run is expected to be similar to that at LEP.

\begin{table}[hbt]
\begin{center}
\vspace*{-0.1in}
\caption{Lower limits at the 95\% CL on the effective Planck scale, $M_S{\rm Hewett}$, in TeV, from HERA. The limits have been translated into Hewett's formalism~\protect\cite{Hewett} from the original formalism~\protect\cite{GRW} used in the H1 and ZEUS analyses.}
\label{H1}
\medskip
\begin{tabular}{|c|cc|cc|cc|}
\hline
Experiment & \multicolumn{2}{c|}{$e^+p$} & \multicolumn{2}{c|}{$e^-p$} & \multicolumn{2}{c|}{Combined} \\
\cline{2-7}
           & $\lambda=+1$ & $\lambda=+1$ & $\lambda=+1$ & $\lambda=+1$ & $\lambda=+1$ & $\lambda=+1$\\
\hline
H1~\cite{H1} & 0.45 & 0.79 & 0.61 & 0.43 & 0.56 & 0.83\\
\hline
ZEUS~\cite{ZEUS} & & & & & 0.66 & 0.66 \\
\hline
\end{tabular}
\end{center}
\end{table}

Based on an analysis of a two-dimensional distribution in the invariant mass and scattering angle of dielectron or diphoton systems, as suggested in Ref.\cite{KCGL}, the D\O\ Collaboration has reported the first search for virtual graviton effects at a hadron collider.\cite{D0} The results, corresponding to 127~pb$^{-1}$ of data collected at $\sqrt{s} = 1.8$~TeV, agree well with SM predictions, and provide the limits on the effective Planck scale shown in Table~\ref{D0}. These limits are similar to and complementary to those from LEP, as different energy regimes are probed at the two colliders. The results of a similar analysis in the dielectron channel from the CDF Collaboration have recently become available,\cite{Gerdes} and are also shown in Table~\ref{D0}. As the current Tevatron sensitivity is limited by statistics, rather than machine energy, we expect the combined Tevatron limits to yield an improvement over the currently excluded range of $M_S$.

\begin{table}[htb]
\begin{center}
\vspace*{-0.1in}
\caption{Lower limits at 95\% CL on the effective Planck scale $M_S$ in TeV, from the Tevatron.}
\label{D0}
\medskip
\begin{tabular}{|c|c|@{}cccccc|@{}cc|}
\hline
Experiment & GRW~\cite{GRW} & \multicolumn{6}{@{}c|}{HLZ~\cite{HLZ}} & \multicolumn{2}{@{}c|}{Hewett~\cite{Hewett}} \\
\cline{3-10}
& & ~~$n$=2 & $n$=3 & $n$=4 & $n$=5 & $n$=6 & $n$=7~~ & ~~$\lambda=+1$ & $\lambda=-1$ \\
\hline
D\O~\cite{D0} & 1.21 & ~~1.37 & 1.44 & 1.21 & 1.10 & 1.02 & 0.97~~  & ~~1.08      & 1.01 \\
\hline
CDF~\cite{Gerdes} & 0.96 &   N/A  & 1.14 & 0.96 & 0.87 & 0.81 & 0.76~~  & ~~0.86      & 0.84\\
\hline
\end{tabular}
\end{center}
\end{table}

Recent attempts to embed LED into string theory,\cite{strings} which predict modified difermion or diboson production at hadron or lepton colliders, as well as phenomenology of Randall-Sundrum~\cite{RS} localized gravity are also being explored. Another exciting topic for future colliders is the possibility to produce black holes, if the collision energy exceeds the effective Planck scale. The detailed phenomenology of black-hole production and decay is currently being developed.\cite{BH}

Although no evidence for LED has been found so far, we are looking forward to the next generation of collider experiments to shed more light on the mystery of extra dimensions. The sensitivity in the $M_S$ reach of the upgraded Tevatron experiments in the next run is expected to double (2~fb$^{-1}$) or even triple (15~fb$^{-1}$), which offers a unique opportunity to see LED effects in the next 5 years. The ultimate test of the theory of large extra dimensions will become possible at the LHC, where effective Planck scales as high as 10~TeV will be able to be probed.


\section*{References}
\begin{thebibliography}{20}
\bibitem{ADD}
N.~Arkani-Hamed, S.~Dimopoulos, and G.~Dvali, Phys. Lett. B{\bf 429}, 263 (1998);
I.~Antoniadis, N.~Arkani-Hamed, S.~Dimopoulos, and G.~Dvali, Phys.Lett. B{\bf 436}, 257 (1998).
\bibitem{tabletop}
C.D.~Hoyle {\it et al.\/}, Phys. Rev. Lett. {\bf 86}, 1418 (2001).
\bibitem{cosmology}
S.~Cullen and M.~Perelstein, Phys. Rev. Lett. {\bf 83}, 268 (1999);
L.~Hall and D.~Smith, Phys. Rev. D {\bf 60}, 085008 (1999);
C.~Hanhart {\it et al.\/}, Nucl. Phys. {\bf B595}, 335 (2001);
S.~Hannestad and G.~Raffelt, e-print hep-ph/0103201 (2001).
\bibitem{GRW}
G.~Giudice, R.~Rattazzi, and J.~Wells, Nucl. Phys. {\bf B544}, 3 (1999) and revised version 2, e-print hep-ph/9811291.
\bibitem{HLZ}
T.~Han, J.D.~Lykken, and R.-J.~Zhang, Phys. Rev. D {\bf 59}, 105006 (1999) and revised version 4, e-print hep-ph/9811350.
\bibitem{Peskin}
E.A.~Mirabelli, M.~Perelstein, and M.E.~Peskin, Phys. Rev. Lett. {\bf 82}, 2236 (1999).
\bibitem{Hewett}
J.L.~Hewett, Phys. Rev. Lett. {\bf 82}, 4765 (1999).
\bibitem{Shrock}
S.~Nussinov and R.~Shrock, Phys. Rev. D {\bf 59}, 105002 (1999).
\bibitem{KCGL}
K.~Cheung and G.~Landsberg, Phys. Rev. D {\bf 62}, 076003 (2000).
\bibitem{common}
R.~Barate {\it et al.\/} (ALEPH Collaboration), ALEPH Note CONF-99-027; {\it ibid.\/} 2000-008;
M.~Acciarri {\it et al.\/} (L3 Collaboration), Phys. Lett. B{\bf 470}, 268 (1999).
\bibitem{direct}
R.~Barate {\it et al.\/} (ALEPH Collaboration), ALEPH Note 2001-010;
P.~Abreu {\it et al.\/} (DELPHI Collaboration), Eur. Phys. J. C{\bf 17}, 53 (2000); DELPHI-CONF-452 (2001); 
G.~Abbiendi {\it et al.\/} (OPAL Collaboration), Eur. Phys. J. C{\bf 18}, 253 (2000).
\bibitem{L3ZZold}
M.~Acciarri {\it et al.\/} (L3 Collaboration), Phys. Lett. B{\bf 464}, 135; {\it ibid.\/} {\bf 470}, 281 (1999).
\bibitem{virtual}
R.~Barate {\it et al.\/} (ALEPH Collaboration), ALEPH Notes 2000-005, 2000-030, 2000-047 (2000);
P.~Abreu {\it et al.\/} (DELPHI Collaboration), Phys. Lett. B{\bf 485}, 45 (2000); {\it ibid.\/} B{\bf 491}, 67 (2000); DELPHI Notes CONF 355, 363, 427, 430 (2000); {\it ibid.\/} 464 (2001);
M.~Acciarri {\it et al.\/} (L3 Collaboration), L3 Notes 2647, 2648 (2001);
G.~Abbiendi {\it et al.\/} (OPAL Collaboration), Phys. Lett. B{\bf 465}, 303 (1999); Eur. Phys. J. C{\bf 13}, 553 (2000); {\it ibid.\/} C{\bf 18}, 253 (2000); OPAL Notes PN 469, 471 (2001).
\bibitem{L3ZZ}
M.~Acciarri {\it et al.\/} (L3 Collaboration), L3 Note 2579 (2000); {\it ibid.\/} 2590 (2000).
\bibitem{OPALZZ}
G.~Abbiendi {\it et al.\/} (OPAL Collaboration), OPAL Note PN 440 (2000).
\bibitem{AD}
K.~Agashe and N.G.~Deshpande, Phys. Lett. B{\bf 456}, 60 (1999).
\bibitem{H1}
C.~Adloff {\it et al.\/} (H1 Collaboration), Phys. Lett. B{\bf 479}, 358 (2000).
\bibitem{ZEUS}
A.~Zarnecki, in Proc. Conf. on Higgs and Supersymmetry, Orsay, March 19-21, 2001.
\bibitem{D0}
B.~Abbott {\it et al.\/} (D\O\ Collaboration), Phys. Rev. Lett. {\bf 86}, 1156 (2001).
\bibitem{Gerdes}
J.~Carlson, talk at the APS Meeting, Washington, D.C., April 28~-- May 1, 2001.
\bibitem{strings}
G.~Shiu, R.~Shrock, and S.-H.~Tye, Phys. Lett. B{\bf 458}, 274 (199); S.~Cullen, M.~Peskin, and M.~Perelstein, Phys. Rev. D{\bf 62}, 055012 (2000).
\bibitem{RS}
L.~Randall and R.~Sundrum, Phys. Rev. Lett. {\bf 83}, 3370 (1999); {\it ibid.\/} 4690 (1999).
\bibitem{BH}
G.~Dvali, G.~Landsberg, and K.~Matchev, paper in preparation.
\end{thebibliography}
\end{document}